\documentclass[aps,prab,reprint,floatfix]{revtex4-2}

\usepackage{graphicx}
\usepackage{siunitx}
\usepackage[caption=false]{subfig}
\usepackage[acronym]{glossaries}

\newacronym{eo}{EO}{electro-optical}
\newacronym{fccee}{FCC-ee}{future circular electron-positron collider}
\newacronym{eosd}{EOSD}{electro-optical spectral decoding}
\newacronym{eos}{EOS}{electro-optical sampling}
\newacronym{pd}{PD}{photodiode}
\newacronym{pbs}{PBS}{polarizing beam splitter}
\newacronym{otr}{OTR}{optical transition radiation}
\newacronym{clear}{CLEAR}{CERN Linear Electron Accelerator for Research}
\newacronym{vm}{VM}{vector modulator}
\newacronym{kara}{KARA}{Karlsruhe Research Accelerator}
\newacronym{fft}{FFT}{Fast Fourier Transform}
\newacronym{cw}{CW}{continous wave}

\begin{document}

\title{An electro-optical bunch profile monitor for FCC-ee}

\author{M. Reissig}
 \email[]{micha.reissig@kit.edu}
\author{E. Bründermann}
\author{S. Funkner}
\author{L.L. Grimm}
\author{B. Härer}
\author{G. Niehues}
\author{J. L. Steinmann}
\author{A.-S. Müller}
\affiliation{Karlsruhe Institute of Technology (KIT), 76131 Karlsruhe, Germany}

\author{R. Corsini}
\author{A. Gilardi}
\author{S. Mazzoni}
 \thanks{Deceased.}
\author{C. Pakuza}
\author{A. Schloegelhofer}
\author{T. Lefevre}
\affiliation{CERN, 1211 Geneva 23, Switzerland}

\author{P. Korysko}
\affiliation{University of Oxford, Oxford, England}
\affiliation{CERN, 1211 Geneva 23, Switzerland}
\date{\today}

\begin{abstract}

	The Future Circular Lepton Collider (FCC-ee) presents challenges for a
	longitudinal bunch profile monitor due to its wide range of bunch lengths
	and charge densities across its four distinct operational modes. For
	commissioning, monitoring the top-up injection, and energy calibration, the
	FCC-ee requires non-destructive, single-shot measurements of the bunch
	length and profile. This contribution proposes an in-vacuum electro-optical
	(EO) longitudinal bunch profile monitor for single-shot measurements at
	high repetition rates, building on the successful EO monitor at the
	Karlsruhe Research Accelerator (KARA) at the Karlsruhe Institute of
	Technology. A novel single-pass conceptual design for the in-vacuum holder
	of the electro-optical crystal is presented, utilizing prisms instead of a
	mirror to guide the laser through the crystal, which additionally allows
	measurements of the long bunches foreseen for FCC-ee operation mode at the
	Z-pole energy. A first prototype has been constructed and tested at the
	in-air test stand of the CERN Linear Electron Accelerator for Research
	(CLEAR). Results from the prototype tests are presented, demonstrating the
	proof of principle for the single-pass prism-based EO monitor design for
	FCC-ee. 

\end{abstract}

\maketitle

\section{Introduction \label{sec:Introduction}}

The \gls{fccee} is planned to be a precision instrument with highest luminosity
to study Z, W, Higgs, and top particles at the smallest
scales~\cite{abada2019}. The FCC innovation study (FCCIS) evaluates the
feasibility of this project in great detail including civil engineering
considerations, beam optics, design of the accelerating RF cavities and beam
guiding magnets as well as beam instrumentation for the commissioning and
operation of the machine~\cite{fccis_webpage}, resulting in the recently
published feasibility report~\cite{benedikt_future_2025_1,
benedikt_future_2025_2}. Within this project, this contribution summarizes the
study of an \gls{eo} bunch profile monitor for \gls{fccee}, including a first
prototype test.

This bunch profile monitor needs to perform single-shot measurements of
high-intensity, picosecond-long particle bunches with the bunch length data
available in a timescale of minutes while still maintaining a sub-picosecond
resolution~\cite[p.~426]{abada2019}. A typically used streak camera for
determining the length of optical pulses is challenging here, as a very
extensive optical beamline would have to be designed due to the large dipole
bending radius of $\sim \qty{9.9}{km}$~\cite{keintzel_fcc-ee_2022}. Another
candidate for longitudinal beam instrumentation is the usage of Cherenkov
diffraction radiation in a dielectric material, which is a newer technique for
non-invasive bunch length measurements and is also under investigation for
\gls{fccee}~\cite{schlogelhofer2024}.

This contribution focuses on an \gls{eo} setup, which makes use of the Pockels
effect in \gls{eo} crystals to probe the Coulomb field of the particle bunches
with laser pulses. This principle was successfully used in an \gls{eosd} setup
for single-shot bunch profile measurements with sub-picosecond resolution at
linear accelerators~\cite{wilke2002,steffen2009,steffen2020}. The \gls{eosd}
setup can also be adapted to circular accelerators, which pose a challenge due
to the MHz-revolution frequencies and wakefield-induced heating, as proven for
the first time by the \gls{kara} at the Karlsruhe Institute of
Technology~\cite{hiller2011,hiller2013}. Its implementation of \gls{eosd} for
single-shot and turn-by-turn bunch profile measurements enables revealing the
dynamics of the ultra-relativistic electron bunches in a non-equilibrium
state~\cite{funkner2023}. This setup serves as a baseline for the development
of an \gls{eo} bunch profile monitor for \gls{fccee}, where unprecedented beam
parameters introduce new challenges for the monitor design.

\section{Electro-optical techniques}

\begin{figure}[ht] 
    \includegraphics{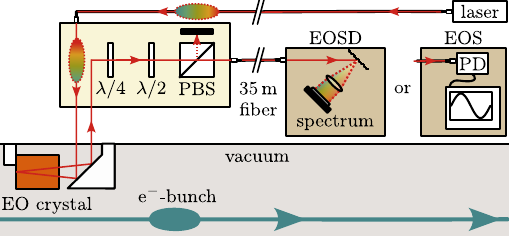}
	\caption{Schematic of electro-optical spectral decoding (EOSD) and
		electro-optical sampling (EOS) at KARA. The laser is guided in a
		\qty{35}{m} long fiber from a laboratory to the vacuum chamber of KARA,
		through the \gls{eo} crystal, a polarizer and back to the laboratory
		for the analysis and data acquisition. The fiber can either be
		connected to a spectrometer for single-shot bunch profile measurements
		with EOSD, or to a photodiode (PD) for an EOS scan of the averaged
		Coulomb- and wakefield of the electron bunch.
		\label{fig:EOSD_schematic}}
\end{figure}

The \gls{eosd} principle is depicted in Fig.~\ref{fig:EOSD_schematic}, which
starts with a chirped laser pulse that is sent into the vacuum chamber and
propagates through an \gls{eo} crystal in parallel to the electron beam. In
this first step, the birefringence of the crystal is modulated by the Coulomb
field of the passing electron bunch due to the Pockels effect. This results in
a polarization modulation of the incident laser pulse while traveling through
the crystal. In the following step, the polarization modulation is translated
to an amplitude modulation by sending it through a \gls{pbs}. The $\lambda/4$
waveplate in front thereby enables compensation for the intrinsic birefringence
of the crystal, while the $\lambda/2$ waveplate is used to set the working
point in a near-crossed configuration~\cite{hiller2013}. In the final step, the
amplitude modulated chirped laser pulse is sent on a spectrometer, consisting
of a grating followed by a focusing lens and a line camera. Due to the linear
chirp, the spectrum corresponds to the longitudinal profile of the laser pulse,
where the Coulomb field of the electron bunch shows as a modulation of the
spectrum. For relativistic particle velocities, the measured Coulomb field
corresponds to the longitudinal charge density profile of the electron bunch
due to the Lorentz contraction, which leads to a longitudinally compressed
field~\cite{ota2022}. For single-shot turn-by-turn measurements, the KIT-built
line camera KALYPSO~\cite{rota2019} is used to enable measurements at the
\gls{kara} revolution frequency of \qty{2.7}{MHz}~\cite{funkner2019}. This
turn-by-turn data has been used to investigate the dynamics of the
micro-bunching instability and to recreate the longitudinal phase space in a
tomographic approach~\cite{funkner2023}. 

For \acrfull{eos} measurements, the spectrometer is replaced by a \gls{pd}
connected to a lock-in amplifier and oscilloscope. Instead of resolving
single-shot laser pulses, the delay of the laser pulse is changed in small
steps, while the recorded \acrlong{pd} signal amplitude provides an indication
of the total laser pulse intensity. The resulting plot of the \acrlong{pd}
signal amplitude over the laser pulse delay provides the timing of the overlap
with the electron bunch, and the general structure of the wakefield.

The achievable resolution is limited by the geometric mean of the
transform-limited laser pulse $\tau_{\text{trans-lim}}$ and the stretched pulse
$\tau_{\text{stretched}}$ with~\cite{sun1998}
\begin{equation} 
\tau = \sqrt{\tau_{\text{trans-lim}} \cdot \tau_{\text{stretched}}} \;. 
\end{equation}

To improve the resolution beyond this limit, methods like electro-optic
spectral interferometry (EOSI)~\cite{walshdavid2023} and phase diversity
electro-optic sampling (DEOS)~\cite{roussel2022} are under development. Since
adjusting \gls{eosd} to either of these methods is possible with minor hardware
changes, this investigation of an \gls{eo} setup for \gls{fccee} focuses on the
more established \gls{eosd} setup.

\section{Simulation of EO measurements under FCC-ee conditions}

The first step toward the development of an \gls{eo} bunch profile monitor for
the \gls{fccee} is to test an established setup under \gls{fccee} conditions to
uncover the challenges that need to be addressed. Since no existing accelerator
can replicate the beam parameters of \gls{fccee}, a simulation procedure for
\gls{eos} measurements has been set up, and its results have been compared to
\gls{kara} measurements to assess its accuracy.

\subsection{Simulation procedure \label{subsec:simulation_procedure}}

The simulations are based on the Wakefield Solver of the CST Particle
Suite~\cite{dassaultsystems2021}, which is used to simulate the electrical
field on the inside of the \gls{eo} crystal using a simplified 3-dimensional
model of the vacuum chamber with the crystal and its holder. The electrical
field is measured using multiple virtual probe points along a line through the
crystal center. Based on this simulation, the total phase retardation caused by
the Pockels effect is estimated in first order by
\begin{equation} \label{eq:phase_retardation} 
\Gamma \approx \sum_i^n \frac{2 \pi d_i }{\lambda} n_0^3 r_{41} E_i^{(y)}(t_i), 
\end{equation}
considering the distance $d_i$ between the probes $i$ and $i+1$, the central
laser wavelength $\lambda \approx \qty{1030}{nm}$ and the vertical part of the
electrical field $E_i^{(y)}$ at probe $i$, as well as the refractive index
$n_0$ of the crystal and its \acrlong{eo} coefficient
$r_{41}$~\cite{steffen2007}. This assumes a crystal orientation for maximum
phase  retardation. For typical \gls{eo} crystal materials like GaP and ZnTe,
this is the case for the Coulomb field and the laser polarization being
parallel to the [-1,1,0] direction of the crystal. The propagation of the laser
also needs to be taken into consideration, which leads to using the probed
field at the arrival time of the laser $t_i$ at probe $i$. The arrival time is
calculated by assuming a constant refractive index $n_0$ at the central laser
wavelength. Due to limitations of the simulation,
Eq.~\ref{eq:phase_retardation} assumes a constant field between the probes,
which is an approximation for small probe spacing. The phase retardation
$\Gamma$ is used in the following sections as a measure to compare the expected
signal strength in different scenarios.

\begin{figure}[ht] 
    \centering
	\includegraphics{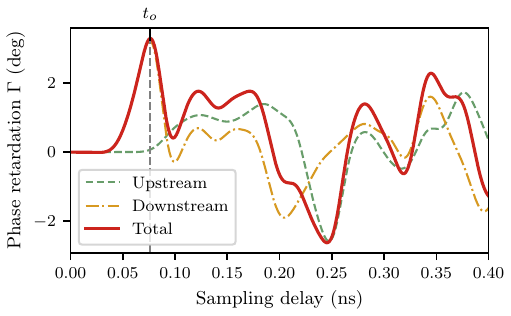}
	\caption{Simulation of the total phase retardation $\Gamma$ at KARA in red,
		with the upstream component in green (dashed) and downstream component
		in yellow (dash-dotted). $t_\text{o}$  highlights the time of the
		Coulomb peak from the bunch overlapping with the laser. The later peaks
		of the phase retardation originate in the wakefield. Around the time of
		the overlap $t_{\text{o}}$, the unwanted upstream signal has a low
	amplitude. \label{fig:up-downstream_vs_sum}} 
 \end{figure}

In the \gls{kara} \gls{eo} setup, the laser initially travels through the
crystal in the upstream direction, opposite to the beam direction. A reflective
coating on backside of the crystal leads to a subsequent propagation in the
downstream direction, in alignment with the beam direction. This two-pass
process is illustrated in the schematic in Fig.~\ref{fig:EOSD_schematic}. In
the simulation, the up- and downstream laser propagation is calculated
separately using Eq.~\ref{eq:phase_retardation}, where the downstream signal
calculation uses the same field probes as upstream, but in reverse order and at
later times.

Thus, the total phase retardation of the laser is the sum of both the upstream
and downstream components, as depicted in Fig.~\ref{fig:up-downstream_vs_sum}
for typical \gls{kara} parameters (for further details of simulation
parameters, see Section~\ref{sec:Sim_vs_Exp}). The shape of the first peak at
$t_{\text{o}}$ in the downstream signal correlates with the Coulomb field of
the electron bunch and is crucial for the single-shot \gls{eosd} measurements.
Although the overlap with the upstream signal can cause disturbances, the
influence on the bunch profile measurement is minimal due to the relatively
small amplitude of the upstream phase retardation around $t_{\text{o}}$.

After the crystal, the laser is guided through a $\lambda/4$-waveplate at angle
$\phi$  followed by a $\lambda/2$-waveplate at angle $\theta$ and a \gls{pbs}.
The angles are defined to be \qty{0}{\degree} for a crossed configuration in
which the laser is not transmitted through the \gls{pbs}. The resulting
detected laser intensity $I_{\text{det}}$ and its modulation through the phase
retardation $\Gamma$ is described by~\cite{steffen2007}
\begin{align} \label{eq:laser_intensity} 
	I_{\text{det}} (\theta, \phi, \Gamma) = \frac{I_{\text{laser}}}{2} [ 1  &- \cos(\Gamma - 2 \phi + 4 \theta)
\cos^2(\phi) \nonumber \\
&+ \cos(\Gamma + 2 \phi - 4 \theta) \sin^2(\phi) ], 
\end{align}
with an initial intensity $I_{\text{laser}}$ before the waveplates. For a
typical operation, the $\lambda/4$ waveplate is set to $\phi =
\qty{0}{\degree}$ to only compensate for the intrinsic birefringence of the
\gls{eo} crystal. With this, the equation is reduced to
\begin{equation} \label{eq:laser_intensity:simple} 
I_{\text{det}} (\theta, \phi=0, \Gamma) = \frac{I_\text{laser}}{2} [1 - \cos(\Gamma + 4 \theta)],
\end{equation}
noindent which highlights the proportionality $I_{\text{det}} \propto 1 -
\cos(\Gamma)$. The working point is typically chosen to be in a near-crossed
setting with $\theta \approx \qty{5}{\degree}$ for two main reasons: First, in
order to maximize the signal-to-noise ratio, the relative modulation of the
laser 
\begin{equation}
    M = \frac{I_{\text{det}} (\theta, \phi=0, \Gamma)}{I_{\text{det}}(\theta,
\phi=0, \Gamma=0)}
\end{equation}
should be as large as possible. In practice, the background from imperfect
laser polarization and polarizer, as well as the noise level and response of
the detector needs to be considered. Second, the correlation between the laser
intensity $I_{\text{det}}$ and the phase retardation should be kept
approximately linear to simplify the interpretation of the \gls{eosd} data.
Therefore, the working point should not be too close to the maximum or minimum
of $I_\text{det}$. Due to the complexity of this optimization, the working
point has been determined empirically for the \gls{kara} setup to be at
$\theta_\text{wp} = \qty{4.6}{\degree}$~\cite{hiller2013}.

As a result, during an \gls{eosd} measurement the modulation $M$ is
approximately linearly correlated with the Coulomb field $E_y$ and hence
corresponds to the longitudinal bunch charge density profile.

\begin{figure}[htb] 
    \centering
	\includegraphics{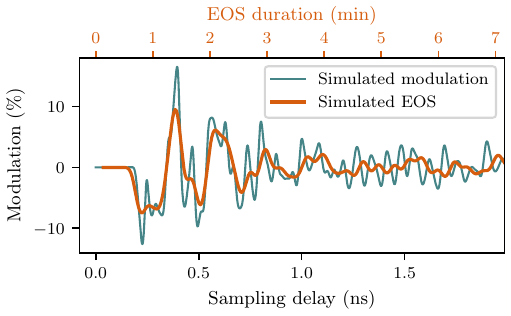}
	\caption{Comparison of the simulated modulation $M$ of the laser pulse
		(blue) and the modulation of a simulated \acrshort{eos} measurement
		(orange) including effects of the laser pulse length, the \acrlong{pd}
		and the lock-in amplifier.\label{fig:ch2:eos_lockin}}
\end{figure}

For the simulation of \gls{eos} measurements, the \acrlong{pd} and the lock-in
amplifier need to be considered. To emulate these devices, it is important to
distinguish between two different time axes:

\begin{itemize}
	\item The sampling delay, which is in the order of nanoseconds and
		describes the relative delay of the laser pulses to the \gls{kara}
		revolution clock. In plots of the laser modulation (e.g.
		Fig.~\ref{fig:ch2:eos_lockin}), the time axis also describes the
		relative arrival time of the Coulomb and wake field.
	\item The measurement duration, which is in the order of minutes and
		describes the time that the performance of an \gls{eos} scan takes. The
		delay of the laser is changed stepwise, and the time for each step is
		typically in the order of seconds.
\end{itemize}

The laser pulses are approximated as a normalized Gaussian with length $\sigma$
multiplied by the modulation $M$. The laser pulse length is generally greater
than the bunch length, because \gls{eosd} measurements require longer pulses to
resolve the bunch profile in single shot measurements. However, for \gls{eos}
measurements, longer pulses limit the resolution, because the lock-in amplifier
output is proportional to the integrated laser pulse intensity. As a result,
the \gls{eos} scan acts similar to a moving average over the laser pulse
length.

The laser pulses are measured with a photodiode, which is approximated as a
low-pass filter, where the cutoff frequency is defined by its bandwidth.
Therefore, this low-pass filter is applied on the sampling delay time.

The lock-in amplifier receives the photodiode signal and a reference signal
with the same repetition rate as the laser, mixes both signals, and applies a
low-pass filter to the output. To highlight signal changes from step to step
during the scan, the time constant of this filter is set slightly shorter than
the time between two EOS scan steps. However, processing every individual pulse
is computationally expensive since single laser pulses are on the picosecond
scale with \unit{MHz} repetition rate, while the duration between EOS scan
steps is in the order of seconds. To streamline the simulation process, two
simplifying assumptions are applied. First, the reference signal multiplication
is omitted, as the simulation does not require artificial noise filtering and
focuses solely on relative modulation. Second, the low-pass filter of the
lock-in amplifier is typically set to have a cut-off frequency of around
\qty{1}{Hz} for \gls{eos} measurements, which only shows long-term changes
during the measurement duration and suppresses the signal of individual laser
pulses. In summary, to emulate the data acquisition, the simulated modulation
is modified by
\begin{enumerate}
	\item multiplying with a normalized gaussian to emulate the laser pulse
		shape for each scan step,
	\item applying a low-pass filter to the emulated laser pulses to simulate
		the \acrlong{pd},
	\item calculating the amplitude maximum for each scan step and adding
		another low-pass filter over the whole scan duration to emulate the
		lock-in amplifier.
\end{enumerate}

A comparison of the simulated modulation $M$ with the simulated relative
modulation after applying the low-pass filters of the \acrlong{pd} and the
lock-in amplifier is presented in Fig.~\ref{fig:ch2:eos_lockin}, using
parameters typical for \gls{eos} measurements at \gls{kara}. These parameters
are a laser pulse length of $\sigma = \qty{20.2}{ps}$,  a \acrlong{pd}
bandwidth of \qty{2}{GHz} and a time constant of $\tau = \qty{1}{s}$ for the
4th order Butterworth low-pass filter of the lock-in assuming a time per
\gls{eos} step of \qty{1.25}{s}. The plot reveals that while some details are
reduced after filtering, \gls{eos} remains valuable to calibrate the timing of
the laser pulses and for qualitative evaluation of the Coulomb and wake-field.
However, high-resolution measurements of the bunch profile need to be done with
a spectrometer using \gls{eosd}.

\begin{figure}[ht]
    \centering
    \includegraphics{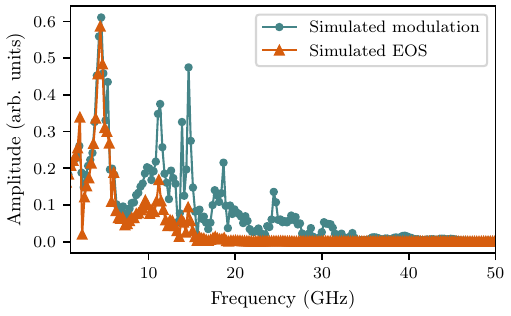}
    \caption{Comparison of the Fourier-transformed simulated modulation $M$
    (blue) and a simulated \gls{eos} measurement (orange), corresponding to of
    the time-domain plots in Fig.~\ref{fig:ch2:eos_lockin}. It shows the
    low-pass filtering of the simulated data acquisition of \gls{eos}
    measurements in comparison to the unfiltered simulated modulation.}
    \label{fig:ch2:eos_lockin_fft}
\end{figure}

A look into the frequency domain is provided in
Fig.~\ref{fig:ch2:eos_lockin_fft} by a \gls{fft} of the simulated modulation
and simulated \gls{eos} measurement. As expected, the low-pass filters to
emulate the photodiode and lock-in amplifier lead to a suppression of higher
frequencies. This is especially noticeable in the region between
\qtyrange{8}{35}{GHz}, where the simulated modulation exhibits multiple
amplitude peaks that are damped in the simulated \gls{eos} measurement.
However, at $f \approx \qty{4.5}{GHz}$ both signals show a prominent peak with
similar amplitude. Additional simulations suggest that this frequency
originates in a resonance of the vacuum chamber, since it changes with the
distance of the crystal to the electron beam and when changing the width of the
vacuum chamber.

\subsection{Comparison of simulations with KARA measurements
\label{sec:Sim_vs_Exp}}

To assess the accuracy of the simulation procedure, an \gls{eos} measurement
was performed at \gls{kara} to detect the electric field of the electron bunch
and its wakefield as a reference~\cite{reissig2024}. While \gls{eosd} enables a
high resolution measurement of the bunch profile with chirped laser pulses over
a time span of tens of picoseconds, \gls{eos} measurements can be used to scan
a longer time span in the nanosecond-scale and beyond around the electron bunch
arrival by changing the delay of the laser pulse and monitoring its intensity
with a \gls{pd}. Therefore, \gls{eos} is useful to find the timing
$t_{\text{o}}$ for the overlap of the Coulomb field of the electron bunch and
the laser pulse inside the crystal. The laser pulse timing can be used for
later \gls{eosd} measurements for longitudinal bunch profile measurements.
However, it can also be used to sample the trailing wakefield of the electron
bunch. This is an important factor for a comparison of experiment and
simulation, since it depends on the geometric impedance of the \gls{eo} crystal
and its holder, which should be similar in both cases.

\begin{figure}[ht] 
    \centering
	\includegraphics{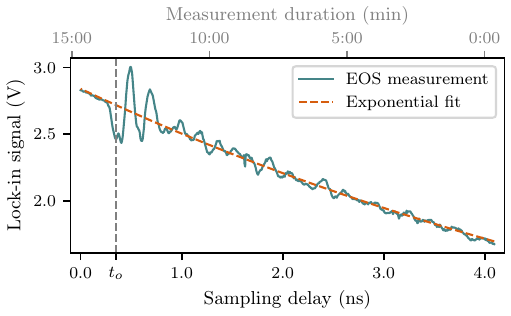}
	\caption{KARA EOS scan with a lock-in amplifier in blue, which
	   corresponds to the measured laser intensity with the
	   \acrlong{pd}~\cite{reissig2024}. The dashed orange line indicates an
	   exponential fit, highlighting the general signal amplitude drift over time.
	   The delay scan started on the right and took \qty{15}{\minute} to complete.
	   \label{fig:KARA_EOS}} 
\end{figure}	

For this measurement, a single bunch was injected into the storage ring, which
was operated at a particle energy of \qty{1.3}{GeV}. The laser repetition rate
is synchronized to the RF signal of \gls{kara} ($\sim\qty{500}{MHz}$) with a
phase locked loop, which enables shifting the laser delay by changing the phase
of the RF signal with a \gls{vm}. For the \gls{eos} scan, the phase was shifted
in steps of $\sim\qty{1}{\degree}$, resulting in a shift of the pulse delay of
$\sim\qty{6}{ps}$ for a total scan range of around \qty{4}{ns}. To improve the
signal-to-noise ratio, the \gls{pd} was connected to a lock-in amplifier using
the \gls{kara} revolution frequency ($\sim \qty{2.7}{MHz}$) as a reference.
Each step of the EOS scan took $\sim\qty{1.25}{s}$, resulting in a total scan
duration of $\sim\qty{15}{min}$. To highlight the signal changes during the
scan, the time constant of the lock-in was set to $\tau = \qty{1}{s}$. 

\begin{figure*}[tbh] 
    \centering
	\includegraphics{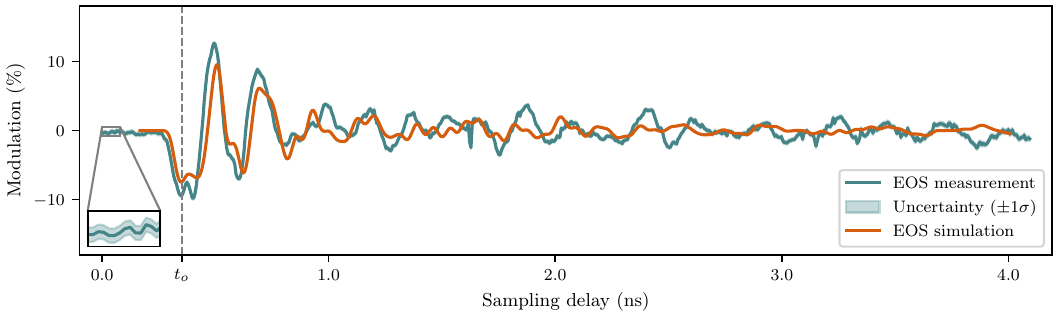}
	\caption{Comparison of an EOS measurement (blue) at KARA with a
	simulation based on CST~\cite{dassaultsystems2021} (orange). The measurement
    includes an error band with $\pm 1 \sigma$ calculated from the signal noise and the
	uncertainty of the baseline fit (Fig.~\ref{fig:KARA_EOS}). The
	simulation corresponds to the \gls{eos} simulation in
	Fig.~\ref{fig:ch2:eos_lockin}. \label{fig:ch2:EOS_Sim_vs_KARA}} 
\end{figure*}

The resulting \gls{eos} scan is presented in Fig.~\ref{fig:KARA_EOS}. The
measurement duration is displayed on the top axis while the delay of the laser
is presented on the bottom axis. The axis of the measurement duration is
reversed, since the scan was done from right to left. During the scan, the
electron bunch charge decreased from approx.~\qty{0.35}{nC} to \qty{0.27}{nC}.
Additionally, the laser intensity slowly increased, likely due to a
beam-induced temperature change of the crystal~\cite{reissig2024}. This changes
the intrinsic birefringence of the crystal and, hence, results in a drift of
the laser polarization. To quantify this drift of the signal baseline, an
exponential function was fitted to the curve. Dividing the lock-in signal by
the exponential fit yields the modulation of the signal.

Figure~\ref{fig:ch2:EOS_Sim_vs_KARA} shows the measured modulation in blue
compared to a simulated \gls{eos} signal in orange. The uncertainty of the
measurement is estimated using the standard deviation of the unmodulated signal
at the sampling delay \qtyrange{0}{0.2}{ns} and the propagated uncertainty of
the parameters from the exponential fit in Fig.~\ref{fig:KARA_EOS}. The total
uncertainty is highlighted in an exemplary magnified inset plot, since the
lock-in amplifier reduces the noise to a total average uncertainty of
$\overline{\sigma}_\text{total} = \qty{0.23}{\percent}$ 

The simulation follows the procedure described in the previous section, but
with waveplate angles of $\phi = \qty{7}{\degree}$ and $\theta =
\qty{2}{\degree}$. The simulation assumes constant waveplate angles, which are
set to the estimated values at the time of the temporal overlap of electron
bunch and laser pulse. Due to the crystal temperature change, the laser
polarization changed slightly over time, which acts like a relative change of
the waveplate angles over time. This effect is described in more detail
in~\cite{reissig2024}. While the simulation assumes a constant bunch charge of
$q(t_\text{o}) = \qty{0.274}{nC}$, the bunch charge decreased during the scan.
Therefore, the measured modulation is scaled with a factor of
$q(t_\text{o})/q(t)$ to compensate the modulation
for the charge loss.

Compared to the measurement, the \gls{eos} simulation in
Fig.~\ref{fig:ch2:EOS_Sim_vs_KARA} closely matches the main features around
$t_{\text{o}}$ at the temporal overlap with the Coulomb field of the electron
bunch but the deviations increase for the peaks with lower amplitude of the
following wakefield. Despite the simplifications made in the simulation, the
amplitude of the main peaks around $t_{\text{o}}$ are close to the measurement.
The modulation amplitude at $t_0$ is $\sim \qty{2.3}{\percent}$ smaller in the
simulation, which could be caused by deviations of the laser polarization or
errors in the estimation of the distance between laser pulse and electron beam,
following Eq.~\ref{eq:laser_intensity} and Eq.~\ref{eq:phase_retardation}. The
increased amplitude of the wakefield in the measured data after around
$t=\qty{1.5}{ns}$ is likely caused by slight changes of the laser polarization
and additional impedances of the vacuum chamber. The simulation only accounts
for a simplified geometry of a short section of the vacuum chamber including
the crystal and its holder. Therefore, additional impedances from other
hardware in and around the vacuum chamber are to be expected. 

\begin{figure}[ht]
    \centering
    \includegraphics{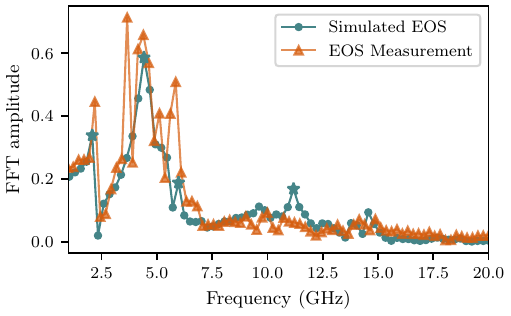}
    \caption{\Gls{fft} of \gls{eos} measurement (blue) and \gls{eos} simulation
    data (orange) from Fig.~\ref{fig:ch2:EOS_Sim_vs_KARA}. It shows a good agreement
    at the most prominent peak at $\sim\qty{4.5}{GHz}$, but it also reveals additional
    peaks that are only present in the measurement. The missing peaks in the
    simulation are likely caused by the simplified geometry of the vacuum
    chamber and the crystal holder.}
    \label{fig:ch2:EOS_Sim_vs_KARA_FFT}
\end{figure}

This assumption is also evident in the frequency domain in
Fig.~\ref{fig:ch2:EOS_Sim_vs_KARA_FFT}, which presents the \gls{fft} of the
simulated and measured \gls{eos} data. While the most prominent peak at $f
\approx \qty{4.5}{GHz}$ is in good agreement, additional peaks for example at
\qtylist{\sim3.7; \sim5.1; \sim5.9}{GHz} do not show in the simulation. As
discussed before, these are probably due to additional impedances caused by
other hardware, which is not represented in the simulation, or caused by the
simplified geometry of the vacuum chamber and crystal holder in the simulation.
However, only the simulation data shows a small peak at \qty{11.2}{GHz}, which
could also be caused by a slightly different geometry or by underestimating the
low-pass filtering in the measurement, for example, by not including the cables
that carry the signal from the photodiode to the lock-in amplifier.

All in all, the simulation procedure can be used to estimate the modulation
amplitude of the Coulomb field and the main features of the wakefield.  The
closely matching wakefield enables to estimate the impedance, which is
discussed later in more detail. These are important factors for the development
of an \gls{eo} setup for \gls{fccee} and can help to identify the challenges of
the demanding beam parameters of the \gls{fccee} operation modes.

\subsection{Simulations under FCC-ee conditions}

With the goal to identify the challenges for an \gls{eo} monitor, the
\gls{kara} \gls{eo} simulation is tested under \gls{fccee} beam conditions.
Since the \gls{fccee} study is still ongoing and the design details are under
investigation, different sets of beam parameters exist for different scenarios
and are frequently updated. In this contribution, the beam parameters are based
on a layout including four interaction points (4 IP), which is currently the
preferred option~\cite{zimmermann2022}. The \gls{fccee} beam parameters applied
in the following simulations are summarized in
Table~\ref{tab:FCC-ee_parameters} and differ for the four planned operation
modes. However, the largest differences in respect of the \gls{eo} setup from
\gls{fccee} to KARA are during Z-operation due to its long bunches and during
t$\bar{\text{t}}$ operation for its high bunch charge density. As a result,
this contribution focuses on these two modes, since the beam parameters for W
operation and ZH operation are in between those two extremes. 

\begin{table}[ht] 
    \caption{FCC-ee beam parameters for the
    \qty{90.7}{km} circumference ring with $4$ interaction points, based
    on~\cite{zimmermann2022}. \label{tab:FCC-ee_parameters}} 
\begin{ruledtabular}
\begin{tabular}{@{}lcccc@{}} 
    Operation mode               & Z     & W     & ZH & t$\bar{\text{t}}$ \\ \hline
    Beam energy (\unit{GeV})     & 45.6  & 80    & 120 & 183               \\
    Bunches / beam               & 10000 & 880   & 248   & 40   \\
    Bunch charge (\unit{nC})     & 38.73 & 46.41 & 32.57 & 37.82             \\
    Bunch length $\sigma_z$ (\unit{mm}) & 15.4  & 8.0   & 6.0   & 2.7 \\
\end{tabular} 
\end{ruledtabular} 
\end{table}

\begin{figure}[ht]
	\includegraphics{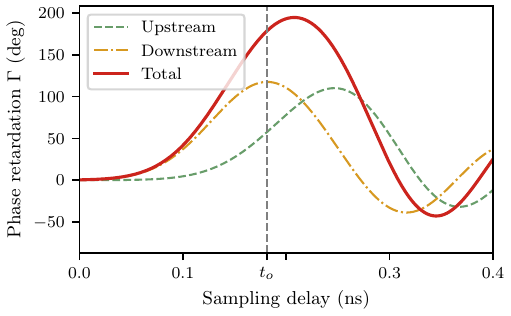}
	\caption{Simulation for the KARA EO design under beam conditions of
	FCC-ee during Z-operation (see Table~\ref{tab:FCC-ee_parameters}).  It shows
	the total phase retardation of the laser in red, with the upstream component
	in green (dashed) and downstream component in yellow (dash-dotted). At  the
	time $t_{\text{o}}$ of the overlap of the Coulomb field with the laser, the
	first peak of the upstream signal overlaps with the downstream signal. This
	would lead to a disturbance of the bunch profile measurement and needs to be
	addressed in a EO setup for FCC-ee. \label{fig:FCC-ee_Z_up-downstream}}
\end{figure}

Figure~\ref{fig:FCC-ee_Z_up-downstream} shows the total phase retardation and
its up- and downstream component for the \gls{kara} \gls{eo} design under
\gls{fccee} Z-mode conditions. The long bunches cause issues with this setup
because, in contrast to the short bunches at \gls{kara}
(Fig.~\ref{fig:up-downstream_vs_sum}), the peak of the upstream signal overlaps
with the peak of the downstream signal on the first peak at $t_{\text{o}}$.
While the phase retardation of the downstream signal correlates directly with
the bunch profile, the correlation to the upstream signal is more complex.
Since only the sum of both can be measured, the upstream signal is considered a
disturbance of the measurement. Additionally, the phase retardation under
\gls{fccee} Z mode conditions peaks at approximately \qty{194.5}{\degree},
whereas the peak at the \gls{kara} simulation has an amplitude of approx.
\qty{3.4}{\degree}. While this can lead to a higher modulation and  better
signal-to-noise ratio, it also causes issues. Preferably, phase retardation is
kept small to achieve an approximately linear relation between the laser
modulation and the Coulomb field strength of the electron bunch (see
Section~\ref{subsec:simulation_procedure} and Eq.~\ref{eq:laser_intensity}). In
addition, the proximity to the high intensity beam leads to high wakefield
intensities and could cause potential damage to the crystal. Due to higher
charge densities during W, ZH and especially t$\bar{\text{t}}$ operation, the
phase retardation is expected to be even higher.

The mitigation of the above-mentioned issues that resulted from the simulations
of the \gls{kara} EO setup under FCC-ee conditions requires a design change of
the setup, which will be addressed in the following sections.

\subsection{Concept idea for FCC-ee}

\begin{figure}[ht] 
    \includegraphics{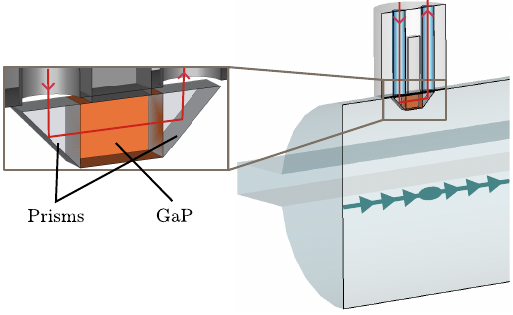}
	\caption{Concept design of a crystal holder for FCC-ee with the GaP
	crystal installed between two prisms. The crystal is attached to the wall of
	the FCC-ee vacuum chamber, which has a diameter of \qty{70}{mm}. The image
	shows a 3D-model created with the CST Studio Suite~\cite{reissig2022a}.
	\label{fig:FCC-ee_Concept_Design}}
\end{figure}

A concept design for an \gls{eo} monitor has been developed, which is adapted
to the beam conditions of \gls{fccee}. Figure~\ref{fig:FCC-ee_Concept_Design}
shows an image of the 3D-model created with the CST Studio
Suite~\cite{dassaultsystems2021}. It includes the planned cylindrical geometry
of the \gls{fccee} vacuum chamber with \qty{70}{mm} diameter and additional
winglets in the horizontal plane~\cite{abada2019}. To avoid the upstream
signal, the laser is guided through the crystal only once in the downstream
direction (single-pass), using prisms attached to its sides. The prisms are
oriented to guide the laser pulses through the crystal using total internal
reflection, such that no reflective coating is necessary. Due to the high
charge density, the distance to the beam can be increased by positioning the
crystal directly below the top of the inner \gls{fccee} vacuum chamber wall.
Additionally, the crystal thickness can be reduced to further decrease the
phase retardation following Eq.~\ref{eq:phase_retardation}. A thinner crystal
also benefits the resolution of \gls{eosd} measurements, since it reduces the
effects of phase mismatch between the laser and the Coulomb field of the
electron bunch.

\begin{figure}[ht]
	\includegraphics{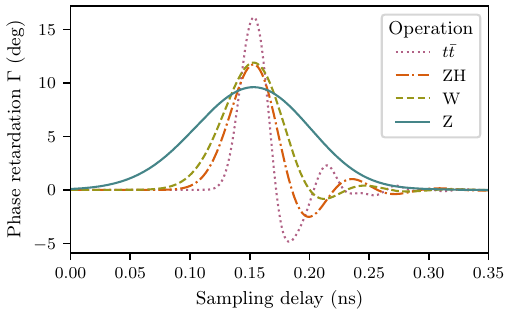}
	\caption{Comparison of the phase retardation $\Gamma$ for different
	FCC-ee operation modes, using varying crystal thicknesses to achieve comparable
	amplitudes. Z (blue, solid) operation is simulated with a \qty{3}{mm} crystal
	thickness, W (green, dashed) with \qty{1}{mm}, ZH (orange, dash-dotted) with
	\qty{1}{mm} and t$\bar{\text{t}}$ (purple, dotted) with \qty{0.5}{mm}.
	\label{fig:Phase-retardation_FCC_modes}} 
\end{figure}

Figure~\ref{fig:Phase-retardation_FCC_modes} displays the phase retardation for
different crystal thicknesses adapted to the FCC-ee operation modes, calculated
from CST simulations. While a \qty{7}{mm} thick GaP crystal is used at
\gls{kara}, it could be reduced to \qty{3}{mm} at \gls{fccee} Z operation.
Despite the smaller crystal, the setup is still achieving a higher phase
retardation of $\Gamma^\text{Z}_\text{max} \approx \qty{9.6}{\degree}$ than the
\gls{kara} setup with $\Gamma^\text{Z}_\text{max} \approx \qty{3.4}{\degree}$.
To achieve a similar amplitude for W and ZH operation, a \qty{1}{mm} crystal is
used in the simulation. t$\bar{\text{t}}$ operation has the largest charge
density in the particle bunches and thus, the crystal thickness is reduced to
\qty{0.5}{mm}, which leads to a maximum phase retardation of
$\Gamma^{\text{t}\bar{\text{t}}}_{\text{max}} \approx \qty{16.1}{\degree}$.

This example shows that the issue of high bunch charge density resulting in a
high phase retardation can be mitigated by placing the crystal on the inner
wall of the vacuum chamber. The increased distance to the particle beam reduces
the phase retardation to a reasonable amplitude, which can be fine-tuned by
adjusting the crystal thickness. Hence, the varying charge density across the
different operation modes of FCC-ee can be accounted for by reducing the
crystal size from Z mode to WW and ZH modes and from ZH to t$\bar{\text{t}}$
mode. Also, the single-pass design of the crystal holder eliminates the
disturbance of the bunch profile measurement by the upstream signal. 

\begin{figure}[ht] 
    \centering 
    \includegraphics{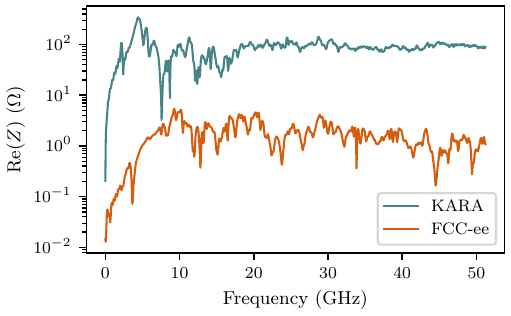}
	\caption{Comparison of the simulated longitudinal impedance Re($Z$)
	at KARA (blue) with a \qty{7}{mm} thick GaP crystal  and the concept for FCC-ee
	(orange) with a \qty{3}{mm} thick crystal. It shows, that the impedance of the
	FCC-cc concepts is generally around 2 orders of magnitude smaller.
	\label{fig:FCC_vs_KARA_impedance}} 
\end{figure}

As an additional benefit of the prism design and the larger distance of the
crystal to the electron beam, the longitudinal impedance is reduced as well.
The simulated real part of the longitudinal impedance is presented in
Fig.~\ref{fig:FCC_vs_KARA_impedance} in comparison to the simulated impedance
at \gls{kara}. While the impedance at \gls{kara} fluctuates around
\qty{100}{\ohm} with the  maximum at $Z^\text{KARA}_\text{max}(\qty{4.4}{GHz})=
\qty{340.8}{\ohm}$, the impedance at \gls{fccee} is around the \qty{1}{\ohm}
mark with the maximum at $Z^\text{FCC-ee}_\text{max}(\qty{9.3}{GHz}) =
\qty{5.4}{\ohm}$. The reduced impedance of the \gls{fccee} concept design is
beneficial to reduce impact on the beam dynamics  as well as reducing the heat
load on the crystal and its holder to minimize a signal drift.

In summary, the simulations show promising results for the concept design for
\gls{fccee}. By attaching the crystal to the wall of the vacuum chamber, the
distance to the beam is large enough to decrease the impedance by an order of
magnitude, while still maintaining a sufficient phase retardation of the laser
pulse. The crystal thickness can be used to adjust the amplitude to the
different operation modes and optimize the resolution. Due to the prisms at the
front and back of the crystal, the laser propagates through the crystal only in
downstream direction and keeps the setup compact. 

\section{Prototype test at CLEAR}

In order to conduct a proof-of-principle experiment, a prototype for the
crystal holder with prisms has been built and tested at the \gls{clear} in
Geneva, Switzerland~\cite{gamba2018}. \gls{clear} is a linear accelerator and
user facility with an in-air test stand at its end. This is the ideal test bed
for a proof-of-principle experiment of the \gls{fccee} \gls{eo} prototype, as
the measurement setup does not need to be vacuum tight and the crystal and
other parts are easily accessible for adjustments. The \gls{clear} electron
beam is created on a CsTe photo-cathode pulsed by an UV (converted from IR)
laser and accelerated by linacs to up to \qty{220}{MeV}. Each electron pulse is
made of 1 to 150 bunches with a bunch separation of \qty{666}{ps}. The pulse
repetition rate ranges from 0.833 to 10 Hz. With bunch charges of up to
\qty{1}{nC} at a bunch length of around \qty{7}{ps},  \gls{clear} provides
suitable parameters for a first test of the \gls{eo} prototype.

\subsection{Prototype design}

\begin{figure}[ht] 
	\centering
	\subfloat[KARA \label{fig:crystal_holder_renders:KARA}]
		{\includegraphics[width=.45\linewidth]{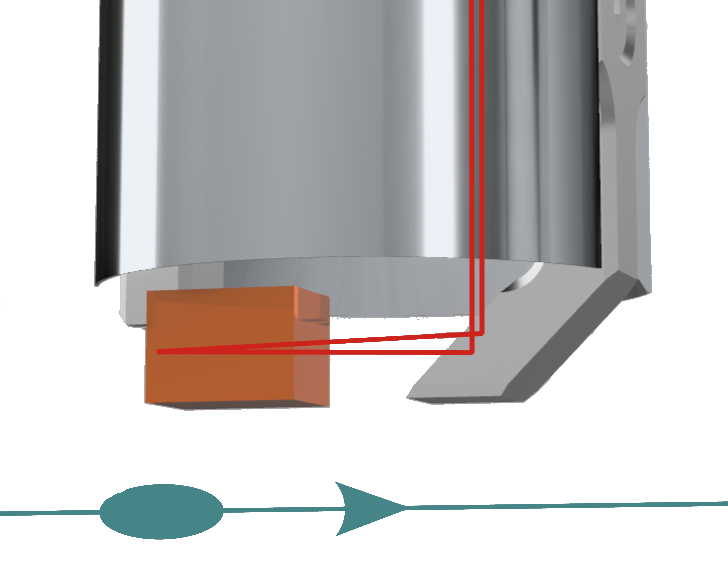}}
	\quad
	\subfloat[FCC-ee prototype \label{fig:crystal_holder_renders:FCCee}]
		{\includegraphics[width=.45\linewidth]{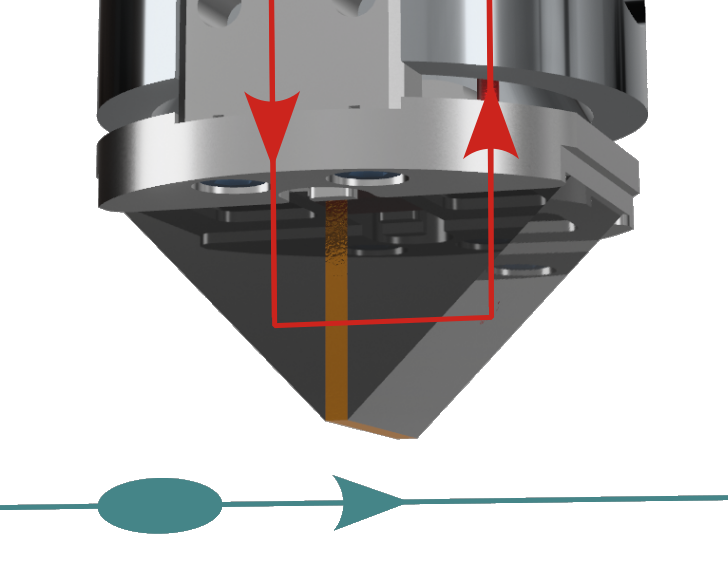}}
	\caption{Rendered 3D models of (a) the crystal holder at KARA and (b)
		FCC-ee prototype for proof-of-principle experiments at CLEAR (CERN).
		The laser path is indicated with a red line and the orbit of the
		particle bunches in blue, while the \gls{eo} crystal is highlighted
		with an orange color. CAD model by S. Schott; rendered by the
		authors.\label{fig:crystal_holder_renders}} 
\end{figure}

The prototype design is based on the crystal holder at \gls{kara}, which is
presented as a rendered image of a 3D-model in
Fig.~\ref{fig:crystal_holder_renders:KARA}. At \gls{kara}, the crystal is glued
to a holder on one side at the end of a metal arm. On the opposite side, a
mirror is attached to guide the laser through the crystal. The crystal has an
anti-reflective coating facing the mirror and a reflective coating on its
backside. As a result, the laser follows down the metal arm onto the mirror,
which guides the laser through the crystal until it hits the reflective
backside. This sends the laser back to the mirror and finally back up through
the metal arm to a metal enclosure, which contains the polarizer setup depicted
in Fig.~\ref{fig:EOSD_schematic}. After the polarizer, the laser is coupled in
a single-mode fiber to eventually end at a \gls{pd}. The crystal used at
\gls{kara} is a GaP crystal with a size of \qtyproduct{5x5x7}{mm}, which is
used in combination with a Ytterbium fiber laser system with a central
wavelength around \qty{1030}{nm}.

The prototype to evaluate the \gls{fccee} concept has a modified crystal
holder, which enables keeping the setup similar to the \gls{kara} version. It
serves as a proof-of-principle for the concept design. A rendered image is
presented in Fig.~\ref{fig:crystal_holder_renders:FCCee}, where the \gls{eo}
crystal is installed between two prisms. Since a metal mirror upstream of the
crystal would shield it from the electric field of the electron bunch, fused
silica prisms are used instead. The prisms are glued to a metal plate at the
top, which has holes to let the laser pass through. Unlike the KARA setup, the
laser is only propagating downstream through the crystal, avoiding the upstream
signal. A \qtyproduct{10x10x2}{mm} ZnTe crystal is used for the \gls{eo}
crystal at \gls{clear} to provide better phase matching with the available
\qty{780}{nm} laser~\cite{wu2014}. 

\subsection{Experimental setup}

\begin{figure}[ht] 
    \centering 
    \includegraphics{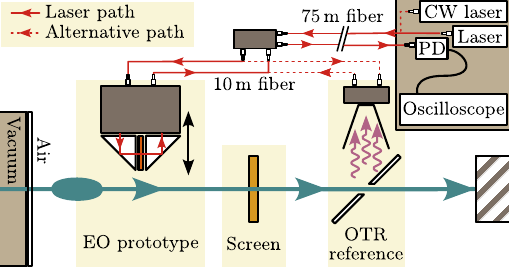}
	\caption{Experimental setup at CLEAR to test the EO prototype for
	FCC-ee.  The laser path is highlighted as a red line, where \qty{75}{m} of
	optical fiber transport the laser light from a pulsed or \gls{cw} laser in a
	laboratory to a fiber optic patch panel at the CLEAR accelerator and back. From
	the patch panel, a \qty{10}{m} fiber can be connected to the \gls{eo} prototype
	or alternatively to an \gls{eo} modulator to detect the \gls{otr} radiation as
	a timing reference.\label{fig:CLEAR_Schematic}}
\end{figure}

Figure~\ref{fig:CLEAR_Schematic} presents a schematic which shows a top-down
view of the in-air experimental area at \gls{clear}. The \gls{eo} prototype is
installed horizontally on a linear stage, which enables adjusting the distance
to the electron beam. A visible alignment laser along the ideal electron beam
path has been used to calibrate the distance between beam center and crystal
during the installation. An additional screen enables alignment of the electron
beam to its center and estimating its horizontal and vertical size.
Additionally, the distance of the crystal can be monitored by looking for an
edge on the screen image, where the crystal starts blocking electrons from
hitting the screen. 

A \qty{780}{nm} pulsed laser with \qty{80}{MHz} repetition rate is delivered
from a nearby laboratory through a \qty{75}{m} long polarization-maintaining
fiber to a patch panel in the experimental area. There, an additional
\qty{10}{m} fiber can be connected  to either the \gls{eo} prototype or an
\gls{eo} modulator to measure the \gls{otr}, which is explained in more detail
in the next paragraph. At the \gls{eo} prototype, the laser is propagating to
the crystal in free-space within a metal enclosure and afterward to motorized
$\lambda/4$ and $\lambda/2$ waveplates followed by a \gls{pbs}. Then, the laser
is coupled back into a \qty{10}{m} + \qty{75}{m} single mode fiber and sent
back to the laboratory. For \gls{eos} measurements, the fiber is connected to a
\gls{pd} with a bandwidth of \qty{12}{GHz}, which is read out by an
oscilloscope with \qty{10}{GHz} bandwidth. However, in order to reduce high
frequency noise from nearby klystrons, the oscilloscope has been digitally
limited to a bandwidth of \qty{3}{GHz}.

\begin{figure}[ht] 
    \centering
	\includegraphics{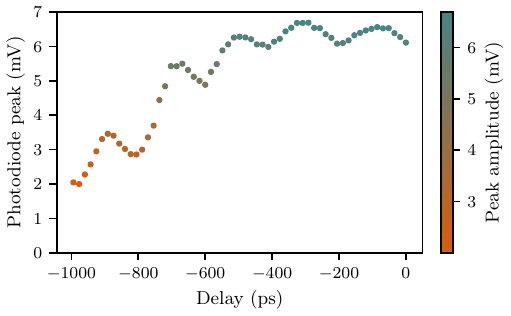}
	\caption{Peak amplitude on the photodiode during a \qty{1}{ns} scan
	with the delay stage. The laser alignment was optimized for the zero-position
	and the amplitude decreases for larger delays due to the movement of the delay
	stage. Each data point shows the average amplitude over \num{100} shots. The
	standard error of the mean is also included, but very small. The delay is
	calculated based on the step size of the linear stage. This color gradient
    is used in the following plots to indicate the peak amplitude.
	\label{fig:EOS_delay-scan_amplitude}}
\end{figure}

\begin{figure}[ht] 
    \centering
    \includegraphics{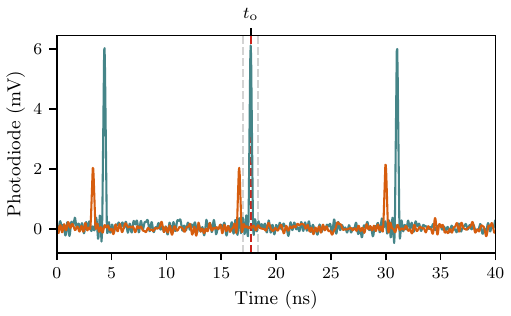} 
    \caption{Oscilloscope trace of the photodiode signal containing
    the peaks of three laser pulses. The blue curve shows the signal at the
    beginning of the delay stage and the orange curve after a delay scan of
    \qty{1}{ns}. The red dashed vertical line describes the estimated position
    $t_{\text{o}}$ for a temporal overlap of the laser pulse with the electron
    bunch. The neighboring two gray lines show the relative position of the
    other two bunches in the train.
    \label{fig:oscilloscope_EOS}} 
\end{figure}

The laser pulse needs to be overlapped in the crystal with the transient
electrical field of the electron bunch. One of the major challenges was
achieving precise timing, as the laser delay could only be adjusted by up to
\qty{1}{ns} using a delay stage. To get a rough estimation of the timing, a
second screen was set up in the beam path to produce \gls{otr}. The \gls{otr}
was collected with a horn antenna and measured with an \acrlong{eo} modulator,
similar to the setup described in~\cite{schlogelhofer2024}. Since the setup
leads to a strong laser modulation, it could be operated with a  \acrlong{cw}
laser, eliminating the need to synchronize the laser pulses with the electron
beam. The \gls{pd} showed the continuous signal of the laser with a peak
occurring every time \gls{otr} was generated by the electron bunch, which was
used as a trigger for the oscilloscope. Since it was using the same fiber
optics as for the FCC-ee EO prototype, the timing of this peak could be  used
as a reference for the \gls{fccee} \gls{eo} prototype with the pulsed laser.

Changing the delay of the laser pulses by moving the delay stage led to slight
misalignment of the laser. This led to changes in laser power at the experiment
due to changes in the coupling efficiency when coupling into the fiber leading
from the laser laboratory to the experiment at \gls{clear}.
Figure~\ref{fig:EOS_delay-scan_amplitude} shows the measured laser amplitude at
the \gls{pd} when changing the delay of the laser pulse by \qty{1}{ns}. The
laser alignment was optimized for the starting point of the delay scan, but
during the scan, the amplitude dropped due to the laser misalignment at the
fiber coupler caused by the movement of the delay stage.

Figure~\ref{fig:oscilloscope_EOS} shows the \gls{pd} signal of three laser
pulses. The blue curve shows the signal at the start of a delay scan, while the
orange curve shows the signal with lower amplitude at the end after a delay
shift of \qty{1}{ns}. Based on the timing of the \gls{otr} reference
measurement and the estimated optical laser path length difference of the EO
prototype, the timing for the temporal overlap with the electron bunch of the
\gls{eo} prototype was estimated and is marked as $t_{\text{o}}$. 

During the measurements, an additional bunch before and after the reference
bunch were produced, to increase the chances of finding the overlap within
range of the delay stage. The expected arrival time of these bunches is
presented as gray dashed lines next to the original bunch at $t_\text{o}$. As a
result, \gls{clear} was producing bunch trains at \qty{10}{Hz}, with each train
consisting of three bunches with a spacing of approx.~\qty{666}{ps} and a bunch
charge of approx.~\qty{1}{nC} per bunch. 

To calculate the modulation, the \gls{pd} signal amplitude needs to be compared
to an unmodulated signal. Therefore, three laser pulses were monitored on the
oscilloscope, but only the second one was expected to overlap with electron
bunches and the other two were used as unmodulated reference.

\begin{figure}[ht]
	\includegraphics[width=.7\columnwidth]{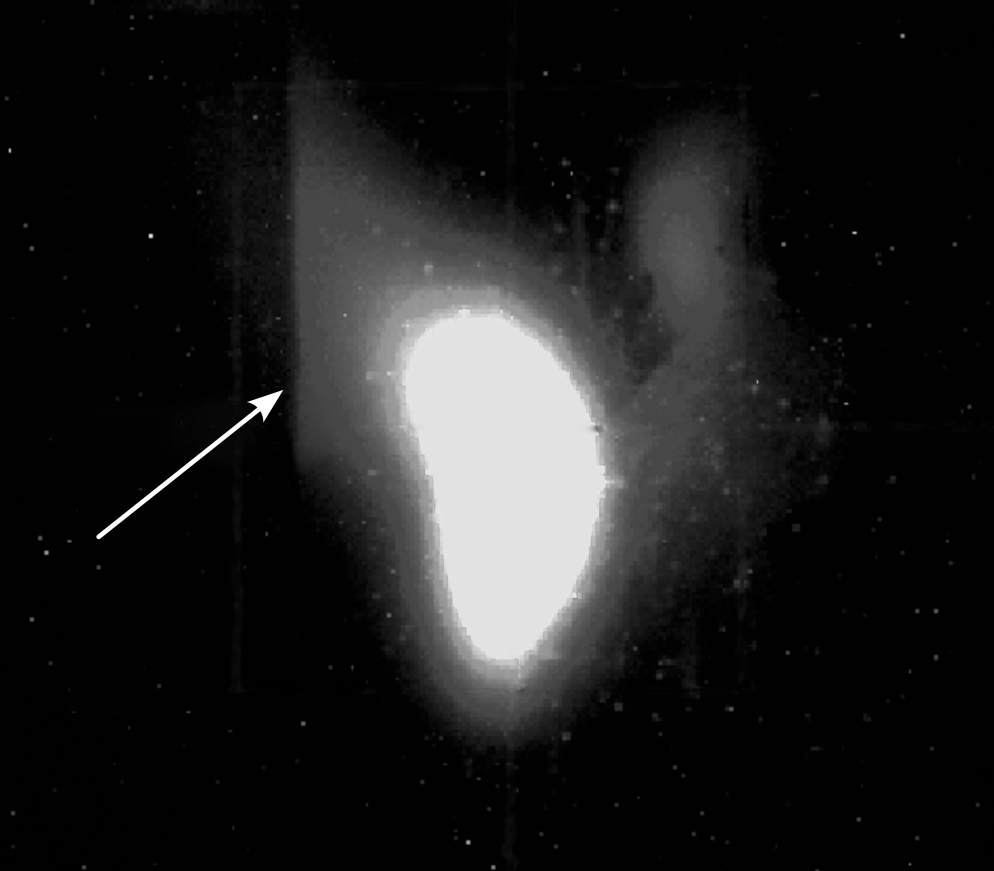}
	\caption{Screenshot of a beam screen image showing the transverse
		shape of the electron beam. A portion of the electron beam is
		obstructed by the EO crystal and prisms, resulting in a subtle shadow
		on the left side of the image, which is highlighted by an arrow. In
		this case, the crystal was moved to an approximate distance of
		$d=\qty{2}{mm}$ to the beam center. The colors have been adjusted and
		transformed to grayscale to improve the visibility of the shadow, while
		the beam center is overexposed. \label{fig:CLEAR_screen}} 
\end{figure}

\subsection{EOS measurements}

To achieve maximum modulation amplitude, the crystal was positioned
approximately $\qty{2}{mm}$ from the electron beam's center. At this distance,
the crystal started blocking some of the outer electrons, which is presented in
the screenshot of a beam screen image in Fig.~\ref{fig:CLEAR_screen}. The
colors have been shifted and transformed to grayscale for better visibility and
a white arrow was added to highlight the crystal shadow. 

For the first test, the working point was set to the angle of the
$\lambda/4$~waveplate to $\phi \approx \qty{0}{\degree}$ and of the
$\lambda/2$~waveplate to $\theta \approx \qty{14.2}{\degree}$. $\theta$ is
chosen to have a greater value than at \gls{kara} to increase the laser signal
amplitude, because, especially toward a laser delay of \qty{1}{ns}, the
signal-to-noise ratio would otherwise be too low.

\begin{figure}[ht] 
	\centering
	\subfloat[$\lambda/2$ waveplate angle $\theta = \qty{+14.2}{\degree}$.
		\label{fig:CLEAR_EOS_modulation:positive}]{
		\includegraphics{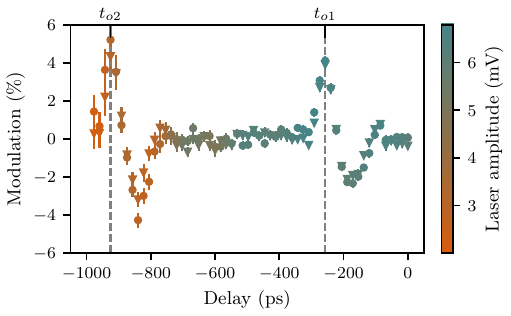}}
	
	\subfloat[$\lambda/2$ waveplate angle $\theta = \qty{-14.2}{\degree}$.
	   \label{fig:CLEAR_EOS_modulation:negative}]{
		\includegraphics{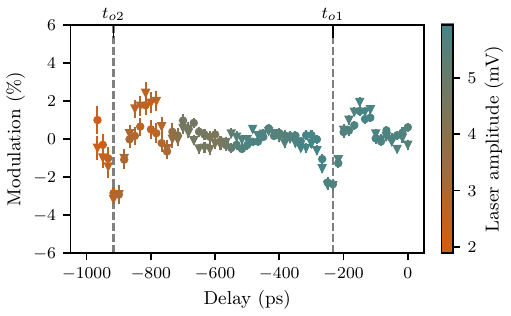}}
	\caption{
		Modulation of laser pulse 2 caused by 2 bunches with a spacing of
		\qty{666}{ps} The circular markers show the modulation of pulse 2
		compared to pulse 3, while the triangular markers show pulse 2 compared
		to pulse 1. The error bars show the standard error of the mean and the
		color gradient indicates the  amplitude of the photodiode signal
		(see~Fig.~\ref{fig:EOS_delay-scan_amplitude}).
    	\label{fig:CLEAR_EOS_modulation}}
\end{figure}

The \gls{eos} scan was done by scanning the delay of the laser pulse to up to
\qty{1}{ns}, while recording \num{100} measurements per step. As described
previously in Fig.~\ref{fig:oscilloscope_EOS}, the oscilloscope monitors three
laser pulses simultaneously, but the modulation by the electron bunch is
expected to only occur at the second bunch. The first and third laser pulse are
used as a reference pulse without modulation. Thus, the circular markers in
Fig.~\ref{fig:CLEAR_EOS_modulation:positive} show the modulation as a division
of the amplitude of the second laser pulse by the third laser pulse, while the
triangular markers shows the division of the second and the first laser pulse.
Every data point and error bar is calculated based on the average of \num{100}
shots and its standard error of the mean, which increases during the scan,
because the laser amplitude decreases (see
Fig.~\ref{fig:EOS_delay-scan_amplitude}). 

The plot shows positive peaks with around \qty{5}{\percent} modulation at
$t_\text{o1}$ and $t_\text{o2}$, which are followed by a ringing. These
positive peaks are \qty{666}{ps} apart, which fits well to the expected bunch
spacing within a bunch train and the ringing could originate in the wakefield.
The laser pulse length was estimated to be a multiple of the electron bunch
length of around $\sigma^{\text{e}^-} = \qty{7.5}{ps}$. Together with the
limited bandwidth of the \gls{pd} and oscilloscope, the \gls{eos} scan was
expected to act similar to a moving average over the actual wakefield and,
therefore, losing the high frequency details (see also
Section~\ref{sec:Sim_vs_Exp}).

This experiment run was followed by another \gls{eos} scan, but with the
$\lambda/2$ waveplate rotated in the other direction to approx. $\theta \approx
\qty{-14.2}{\degree}$ to test, if the \gls{eos} measurement behaves as
expected. If the previously observed modulation is the \gls{eos} signal of an
electron bunch, it should satisfy Eqs.~\ref{eq:laser_intensity} and
\ref{eq:laser_intensity:simple} and lead to the modulation having the opposite
sign. The result is plotted in Fig.~\ref{fig:CLEAR_EOS_modulation:negative},
which shows the expected result of two negative peaks at $t_\text{o1}$ and
$t_\text{o2}$ followed by a ringing. However, the modulation amplitude dropped
to around \qty{2.5}{\percent}, the cause of which is not yet fully understood.
The second measurement was performed a few hours after the first and due to
machine operation issues, the machine parameters needed to be re-adjusted to
optimize the bunch charge transport through CLEAR. The modulation difference is
likely caused by a slightly lower bunch charge compared to the previous
measurement and a potentially high uncertainty on the angle of the
$\lambda/2$~waveplate. 

In summary, these measurements of the first \gls{eo} monitor prototype for
\gls{fccee} demonstrate the viability of using prisms attached to the crystal.
This innovative design enables bunch profile measurements for both long and
short bunches while maintaining a compact geometry, marking a substantial step
forward in the development of a longitudinal bunch diagnostic that meets the
requirements for the various operation modes of \gls{fccee}.

\section{Summary} 
\glsresetall 

The development of advanced longitudinal beam diagnostics is essential for the
future electron-positron collider to monitor bunch length and bunch profile
with high precision. This study demonstrates the feasibility of an \gls{eo}
setup, which specifically addresses the challenge of being able to measure the
profile of long bunches as well as short bunches with high bunch charge. To
assess the specific challenges at the \gls{fccee}, a simulation procedure has
been developed based on the \gls{eo} setup at the \gls{kara}. The setup at
\gls{kara} is optimized for low bunch charges and short bunches and has a
two-pass design, where the laser travels upstream through the crystal, is
reflected at the backside and subsequently travels downstream parallel to the
electron bunch. This design limits its use for longer bunches, because the
upstream signal starts to overlap with the bunch profile in the downstream
signal, making it difficult to extract the bunch profile in the final
measurement. Therefore, it is unsuitable for measuring the long, high-charge
bunches during Z-operation at FCC-ee. Thus, an innovative single-pass design of
the crystal holder has been developed, which uses two prisms attached to the
sides of the crystal to guide the laser through the crystal only once. The
novel design enables measurements of long and short bunches while still keeping
the setup  compact with substantially reduced impedance. A corresponding
prototype has been built for initial testing at the in-air test area at the
\gls{clear}. 

The prototype was successfully tested with \gls{eos} measurements at the in-air
test area at \gls{clear}, yielding a clear scan of two electron bunches and
their wakefield, thereby validating the concept’s functionality. This
proof-of-principle marks a milestone toward realizing a longitudinal bunch
diagnostic, which meets the requirements for \gls{fccee}. Future work will
focus on optimizing the design for installation at \gls{fccee} and \gls{eosd}
tests along with technical details about the laser and crystal material to
enhance the resolution of bunch profile measurements. In addition, measurements
during multi-bunch operation at \gls{fccee} provide challenges which need to be
investigated, especially considering an increased heat load and exposure to
radiation.

\paragraph*{Note added.} While this manuscript was under preparation, one of the
coauthors, Stefano Mazzoni, sadly passed away.

\section{Acknowledgements}

The authors thank Steffen Schott for assistance with the CAD design of the
experimental setup. This project has received funding from the European Union's
Horizon 2020 research and innovation programme under grant agreement No 951754
(FCCIS) and No 101057511 (EURO-LABS). M.~R.~acknowledge funding from the
Bumdesministerium für Bildung und Forschung (BMBF) under contract number
05K22VKB.

\bibliographystyle{apsrev4-2}  
\bibliography{2025_EOforFCCee_arXiv}  

\end{document}